# Free-running single-cavity dual combs with Hz-level relative linewidth

*Yuan Chen[1,2], Ming Yan[1,2]*, Jia Xu[1], Yueting Hu[1], Jieming Zhang[1], Zhaoyang Wen[1,2]*, Zijian Wang[1], Xiangze Ma[1], Min Li[3]*, Heping Zeng[1,2,4]*

Yuan Chen, Ming Yan, Jia Xu, Yueting Hu, Jieming Zhang, Zhaoyang Wen, Zijian Wang, Xiangze Ma, Heping Zeng

State Key Laboratory of Precision Spectroscopy, Hainan Institute, East China Normal University, Shanghai 200062, China

E-mail: zywen@lps.ecnu.edu.cn; myan@lps.ecnu.edu.cn

Yuan Chen, Ming Yan, Zhaoyang Wen, Heping Zeng

Chongqing Key Laboratory of Precision Optics, Chongqing Institute of East China Normal University, Chongqing 401120, China

Min Li

School of Optical-Electrical and Computer Engineering, University of Shanghai for Science and Technology, Shanghai 200093, China

E-mail: minli1220@usst.edu.cn

Heping Zeng

Jinan Institute of Quantum Technology, Jinan, Shandong 250101, China

## Abstract:

Single-cavity dual-comb lasers provide a compact and efficient source for dual-comb spectroscopy in gas sensing applications; however, achieving sufficient free-running mutual coherence for comb-line-resolved, high-resolution measurements remains challenging. Here, we present a symmetry-engineered bidirectional single-cavity dual-comb laser based on an all-polarization-maintaining fiber architecture. The system exhibits exceptional free-running mutual coherence, achieving Hz-level relative linewidths without active feedback or phase correction. The time-averaged absolute jitter of the dual-comb repetition-rate difference reaches $4.7\times10^{-7}$ $min^{-1}$, representing an improvement of nearly two orders of magnitude over previously reported free-running systems. As a spectroscopic demonstration, we resolve ~49,000 comb lines over a 5.4 THz optical bandwidth and measure the absorption spectrum of carbon monoxide ($^{12}CO$), faithfully retrieving molecular line shapes with millisecond acquisition times. This architecture provides a compact and robust free-running platform for broadband molecular spectroscopy and millisecond-scale, line-shape-resolved gas sensing.

## 1. Introduction

Dual-comb spectroscopy (DCS) has emerged as a powerful technique for rapid, high-resolution, high-sensitivity, and broadband spectroscopic measurements by mapping optical spectra directly into the radio-frequency (RF) domain without mechanical scanning [1-6]. Its performance largely relies on maintaining high mutual coherence between the two optical combs, as stable phase relationships ensure narrow comb-line resolution, high signal-to-noise ratio (SNR), and coherent averaging over extended acquisition times [7]. However, preserving such coherence is challenging because fluctuations in the repetition rates and carrier-envelope-offset frequencies of independently mode-locked lasers introduce relative timing and phase noise [8, 9]. These instabilities broaden the RF comb lines and degrade spectral fidelity, generally necessitating sophisticated phase-locking or correction schemes. Consequently, conventional DCS systems based on two independent comb sources often involve complex control architectures, which hinder their deployment in practical measurement environments [10, 11].

To overcome the complexity associated with active stabilization, single-cavity dual-comb (SCDC) systems have emerged as an attractive alternative [12, 13]. By generating two asynchronous pulse trains within a shared laser cavity, these systems inherently suppress common-mode fluctuations and enhance mutual coherence, thereby simplifying the system architecture and reducing the need for elaborate control electronics. So far, various approaches, including wavelength multiplexing [14, 15], polarization multiplexing [16, 17], spatial multiplexing [13], and bidirectional operation [18, 19], have been explored to realize dual-comb generation. Despite these advances, achieving comb-line-resolved spectroscopy directly from a free-running SCDC source remains challenging [20, 21]. Although the two combs share a common cavity, residual non-common perturbations can still arise from intracavity asymmetries, unequal responses of the counter-propagating pulse trains to environmental disturbances, and intracavity gain competition [13, 18]. These effects introduce differential timing and phase fluctuations between the combs, leading to RF comb-line broadening and reduced coherent averaging efficiency. As a result, the mutual coherence of free-running SCDC systems often remains insufficient for high-precision, comb-line-resolved measurement. To mitigate these limitations, adaptive sampling schemes [22] and phase-correction algorithms [20] have been employed to compensate for relative fluctuations between the combs. However, these approaches introduce substantial computational and instrumental complexity, partially offsetting the simplicity advantages of single-cavity architectures.

In this work, we demonstrate a highly stable bidirectional SCDC laser. Central to the design is a symmetry-preserving nonlinear polarization rotation (NPR) mode-locking architecture with polarization-mode-dispersion (PMD) compensation in a polarization-maintaining (PM) fiber. By minimizing differential cavity perturbations, this approach substantially enhances the mutual coherence between the two combs. The free-running system exhibits an Allan deviation of 1.33 mHz at an integration time of 1 s and a standard deviation (SD) of 12.7 mHz over 2 h. More importantly, the relative dual-comb linewidth reaches the Hz level without active stabilization or phase correction. As a spectroscopic demonstration, we resolve approximately 49,000 RF comb lines across a 5.4 THz optical bandwidth and measure the absorption spectrum of carbon monoxide ($^{12}CO$) with faithful recovery of molecular line shapes at millisecond acquisition times. These results establish a compact and robust platform for broadband, comb-line-resolved molecular spectroscopy, paving the way for practical applications in high-accuracy gas sensing and chemical analysis.

## 2. Principle and Experimental Setup

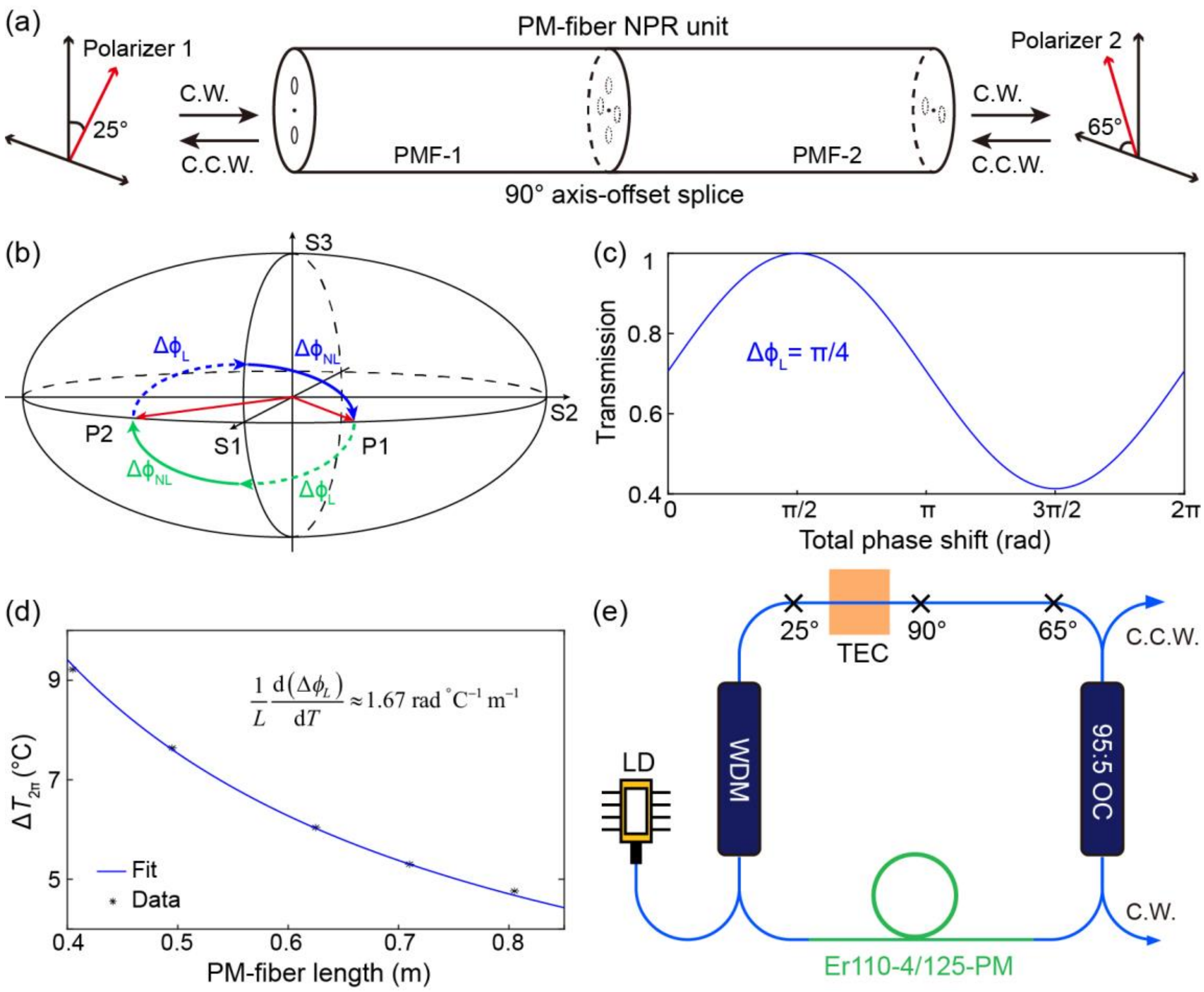

**Figure 1. Principle and implementation of the all-PM NPR bidirectional SCDC laser.** a) Schematic of the PM-fiber NPR unit. Two nearly equal-length PM-fiber segments are spliced with a 90° axis offset to compensate polarization-mode dispersion and suppress Lyot filtering. b) Polarization evolution of the clockwise (C.W.) and counter-clockwise (C.C.W.) pulses on the Poincaré sphere. c) Calculated NPR transmission as a function of the total phase shift for an offset angle of 25°. d) Measured and fitted values of the temperature change $\Delta T_{2\pi}$ corresponding to a $2\pi$ shift in the linear phase difference versus PM-fiber length. e) Experimental setup of the all-PM bidirectional single-cavity dual-comb laser. WDM, wavelength division multiplexer; OC, optical coupler; TEC, thermoelectric cooler; LD, laser diode.

The working principle of the PM-fiber NPR unit is illustrated in Figure 1(a). Two nearly equal-length PM-fiber segments are spliced with a 90° axis offset. This configuration swaps the fast and slow axes between the two segments, thereby compensating the first-order polarization-mode dispersion accumulated in the first segment and suppressing the Lyot-type spectral filtering that would otherwise arise from the strong intrinsic birefringence of PM fibers [23]. This tailored arrangement enables a reproducible NPR evolution process with an all-PM fiber cavity, while preserving near-identical operating symmetry for the two counter-propagating pulse trains.

When linearly polarized light enters a PM fiber with its polarization axis misaligned relative to the fiber's principal axes, the two orthogonal polarization components accumulate both linear and nonlinear phase shifts as they propagate. After passing through an intracavity polarizer, the resulting phase-dependent polarization state is converted into an intensity-dependent transmission. This effect mimics an artificial saturable absorber and forms the basis of NPR mode locking [24]. Owing to the symmetric design of the PM fiber cavity, the clockwise (C.W.) and counter-clockwise (C.C.W.) pulses experience nearly identical NPR evolution. This symmetry is essential for achieving stable bidirectional mode locking in a shared cavity.

To describe this process quantitatively, the evolution of the pulse envelopes along the two principal axes can be written as [25]:

$$\frac{\partial A_x}{\partial z} + \beta_{1x}\frac{\partial A_x}{\partial t} + \frac{i\beta_2}{2}\frac{\partial^2 A_x}{\partial t^2} = i\gamma\left(|A_x|^2 + \frac{2}{3}\left|A_y\right|^2\right)A_x + \frac{i\gamma}{3}A_x^* A_y^2 \exp(-2i\Delta\beta z) \qquad (1)$$

$$\frac{\partial A_y}{\partial z} + \beta_{1y}\frac{\partial A_y}{\partial t} + \frac{i\beta_2}{2}\frac{\partial^2 A_y}{\partial t^2} = i\gamma\left(\left|A_y\right|^2 + \frac{2}{3}|A_x|^2\right)A_y + \frac{i\gamma}{3}A_y^* A_x^2 \exp(2i\Delta\beta z) \qquad (2)$$

$$\Delta\beta = \beta_{0x} - \beta_{0y} = 2\pi/L_B \qquad (3)$$

where $A_x$ and $A_y$ are the pulse envelopes along the two orthogonal principal axes; $\beta_{1x}$ and $\beta_{1y}$ are the first-order dispersion parameters along the x and y axes, respectively; $\beta_2$ is the second-order dispersion parameter assumed to be identical for the two axes; $\gamma$ is the nonlinear coefficient; and $L_B$ is the beat length of the PM fiber. In our PM fiber, $L_B$ is on the order of millimetres, whereas the nonlinear characteristic length ($L_{NL}$) required to accumulate a sub-$\pi$ nonlinear phase shift is on the order of meters, satisfying $L_{NL} \gg L_B$. Consequently, over the propagation distance relevant to nonlinear phase accumulation, the phase mismatch $\Delta\beta = 2\pi/L_B$ causes the coherent coupling terms containing $\exp(\pm 2i\Delta\beta z)$ to oscillate rapidly. These fast oscillations average out to zero, effectively decoupling the two polarization axes [25]. Under this approximation, the PM-NPR unit primarily introduces a controllable phase difference between the two polarization components, consisting of a linear phase delay $\Delta\phi_L$ and an intensity-dependent nonlinear phase shift $\Delta\phi_{NL}$.

This phase difference determines the polarization trajectory on the Poincaré sphere, as shown in Figure 1(b). Subsequently, the projection by polarizers $P_1$ and $P_2$ converts this phase evolution into a variation in transmission. The calculated transmission curve shown in Figure 1(c) presents the dependence of the output transmission on the total phase shift $\Delta\phi_L + \Delta\phi_{NL}$ for a selected axis-offset angle. By choosing an appropriate operating point on this curve, an increase in pulse peak power results in higher transmission, thereby generating the saturable-absorption response based on the NPR effect, which is essential for initiating mode locking.

Specifically, the offset angles, defined as the angle between the slow axes of the two PM fibers, were set to 25° and 65°. This configuration balances low intracavity loss with high modulation depth, a condition proven to be highly conducive to achieving mode locking [26, 27]. The linear phase difference, $\Delta\phi_L$, arises from the imperfect symmetry between the fiber segments on either side of the 90° splice—caused by factors such as length mismatch, temperature, and stress—resulting in incomplete compensation of the PMD. In Figure 1(c), $\Delta\phi_L$ is assumed to be $\pi/4$. To facilitate the thermal tuning of $\Delta\phi_L$, we experimentally characterized the temperature change required to induce a $2\pi$ phase shift for various fiber lengths (Figure 1(d)). For a 40 cm fiber segment, a temperature variation of only 9.2 °C is sufficient to shift $\Delta\phi_L$ by $2\pi$, ensuring arbitrary control of the phase difference near room temperature. Additionally, the thermal sensitivity coefficient for the phase difference per unit length of the PM fiber was determined to be 1.67 rad $℃^{-1}$ $m^{-1}$.

The detailed experimental configuration is shown in Figure 1(e). The cavity employs a wavelength division multiplexer (WDM) and a 95:5 optical coupler (OC), both of which are fast-axis-blocked components that serve as the essential in-cavity polarizers. In the upper section of the resonator, the WDM's pigtail is spliced to a ~40 cm segment of PM fiber at an offset angle of 25°. This segment is then spliced to another ~40 cm PM fiber at 90° to compensate for the PMD between the fast and slow axes. During fabrication, the length difference between these two segments is strictly controlled to within 1 cm, corresponding to a differential group delay of ~10 fs, which is a value that negligibly affects the NPR evolution. To enable arbitrary control of the linear phase difference $\Delta\phi_L$, one of the PMF segments is mounted on a thermoelectric cooler (TEC) for precise thermal management. On the opposite side of the cavity, a 976 nm pump laser with a maximum output power of 700 mW is coupled into the resonator via the WDM, pumping a ~40 cm segment of erbium-doped fiber (EDF, Er110-4/125-PM). Stable bidirectional mode-locking is achieved by optimizing $\Delta\phi_L$.

## 3. Results

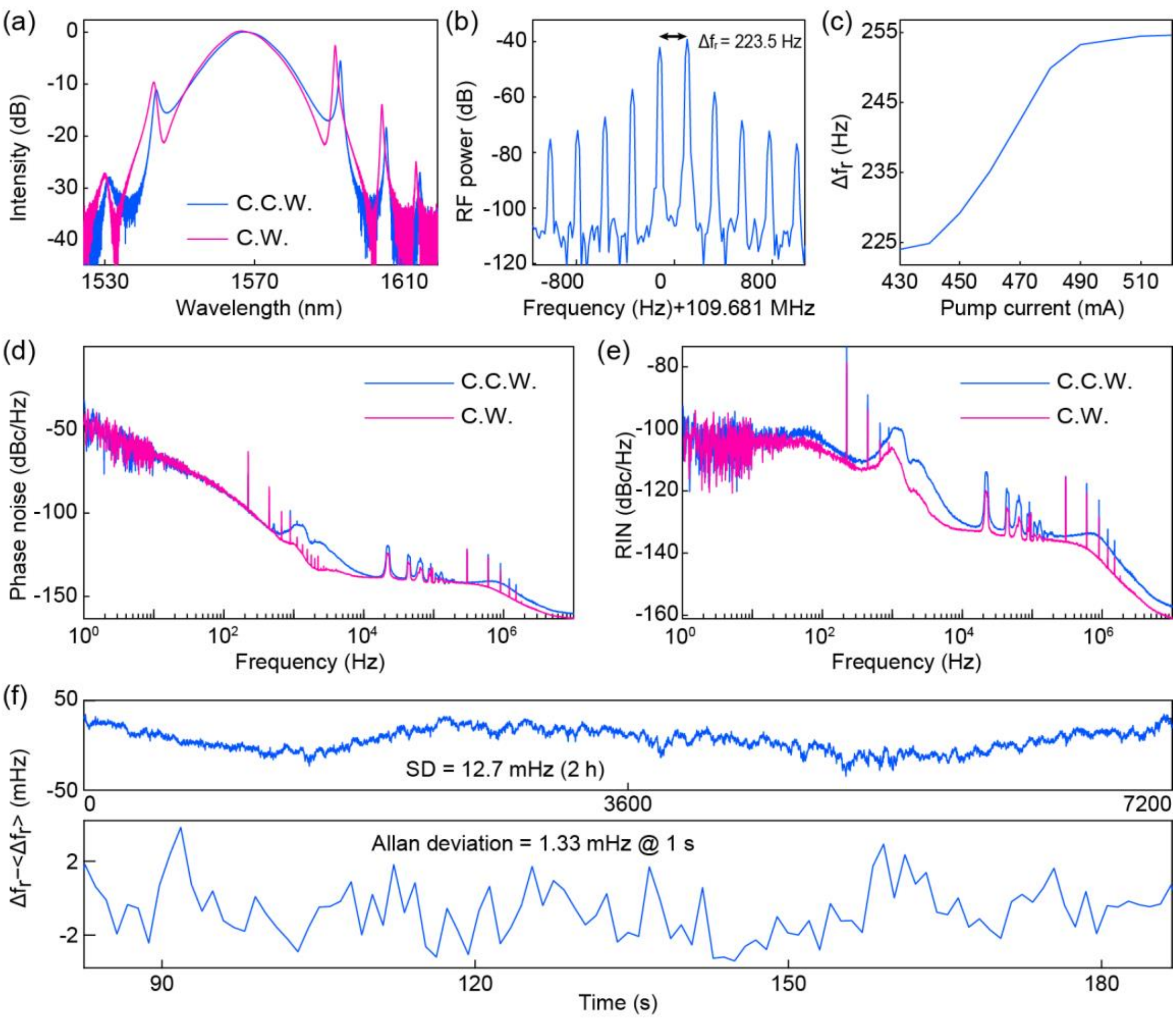

**Figure 2.** Characteristics of the bidirectional SCDC system. a) Optical spectra of the C.W. and C.C.W. pulse trains. b) RF spectrum around the fundamental repetition rate, showing the repetition-rate difference $\Delta f_r$. c) Pump-current tuning of $\Delta f_r$. d,e) Phase-noise and relative intensity noise (RIN) spectra of the two pulse trains, respectively. f) Free-running stability of $\Delta f_r$, with the upper panel showing the 2 h drift and lower panel showing the short-term fluctuation.

Figure 2(a) shows the optical spectra of the C.W. and C.C.W. pulses, whose center wavelengths differ by approximately 0.8 nm. This wavelength offset arises from asymmetric gain and nonlinear interactions experienced by the two counter-propagating pulses and, through cavity dispersion, gives rise to the group-velocity mismatch required to generate a finite repetition-rate difference ($\Delta f_r$) [28]. Despite this offset, the two combs retain substantial spectral overlap, ensuring efficient dual-comb spectroscopic measurements. Furthermore, the two pulse trains exhibit a repetition rate of $f_r$=109.68 MHz and a repetition-rate difference of $\Delta f_r = 223.5$ Hz (Figure 2(b)). Both values are consistent with predictions based on the measured cavity length (~181 cm) and net intracavity dispersion (23 fs nm$^{-1}$). In addition, in our system, $\Delta f_r$ can be continuously tuned by adjusting the pump current (Figure 2(c)). This behavior arises from pump-induced changes in the intracavity gain and nonlinear phase accumulation, which modify the relative group velocities of the counter-propagating pulse trains.

We characterized the phase noise and relative intensity noise (RIN) of the two pulse trains at their fundamental repetition frequencies. As shown in Figure 2(d,e), the C.W. and C.C.W. pulses exhibit similar noise characteristics, reflecting the strong common-mode noise suppression enabled by the shared cavity. At the same time, subtle differences in their spectral profiles are observed. In particular, weak peaks emerge at $\Delta f_r$ and its harmonics, which are attributed to residual coupling between the counter-propagating pulses through cross-gain modulation in the gain fiber and weak intracavity back-reflections [29]. Although these features indicate incomplete isolation of the two pulse trains within the cavity, their amplitudes remain small and do not significantly contribute to the overall phase-noise or RIN performance. The remaining differences between the two directions are likely caused by the distinct sequence in which the pulses experience gain amplification, NPR evolution, and output coupling during each cavity round trip.

**Table 1**. Comparison of free-running bidirectional SCDC systems

| $f_r$ (MHz) | $\Delta f_r$ (Hz) | $\sigma(\Delta f_r)$ (Hz) | $TAA\ jitter$ (min$^{-1}$) | Ref. |
|---|---|---|---|---|
| 27.97 | 116 | 0.78 in 100 min | $6.72\times10^{-5}$ | [30] |
| 45.9 | 140 | 2.5 in 100 min | $1.79\times10^{-4}$ | [31] |
| 11.778 | 202 | 4.26 in 50 min | $4.22\times10^{-4}$ | [32] |
| 9.99 | 224 | 0.1882 in 5 min | $1.68\times10^{-4}$ | [33] |

| 78.04 | 1282 | 0.7 in 20 min | $2.73\times10^{-5}$ | [34] |
|---|---|---|---|---|
| 109.68 | 223.5 | 0.0127 in 120 min | $4.74\times10^{-7}$ | This work |

To evaluate the relative stability between the two counter-propagating combs, we measured the repetition-rate difference, $\Delta f_r$, under free-running operation. As shown in Figure 2(f), $\Delta f_r$ was continuously recorded over 2 h with a 1 s gate time. The measured $\Delta f_r$ shows a standard deviation (SD) of 12.7 mHz over the full measurement period, while the Allan deviation reaches 1.33 mHz at an integration time of 1 s, indicating excellent short- and long-term stability.

Direct comparison of the SD of $\Delta f_r$ among different studies, however, can be misleading because the reported values strongly depend on the observation duration and measurement conditions. To mitigate this issue and enable a more meaningful comparison of free-running stability, we adopt the time-averaged absolute (TAA) jitter proposed in a previous study [35]:

$$TAA\ jitter = \frac{\sigma(\Delta f_r)}{\Delta f_r} \cdot \frac{1}{T_{span}} \tag{4}$$

where $\sigma(\Delta f_r)$ is the standard deviation of the repetition-rate difference, $\Delta f_r$ is the mean repetition-rate difference, and $T_{\text{span}}$ is the total measurement time. By normalizing the relative fluctuation of $\Delta f_r$ to the observation time, this metric provides a more consistent basis for comparing free-running stability across different systems.

Using this metric, we benchmarked our laser against previously reported bidirectional SCDC systems (Table 1). The proposed all-PM NPR cavity achieves a TAA jitter of $4.74 \times 10^{-7}\ \text{min}^{-1}$, representing an improvement of nearly two orders of magnitude over the best previously reported result [34]. These results highlight the effectiveness of the symmetry-preserving all-PM architecture in suppressing differential cavity perturbations and stabilizing free-running bidirectional dual-comb operation.

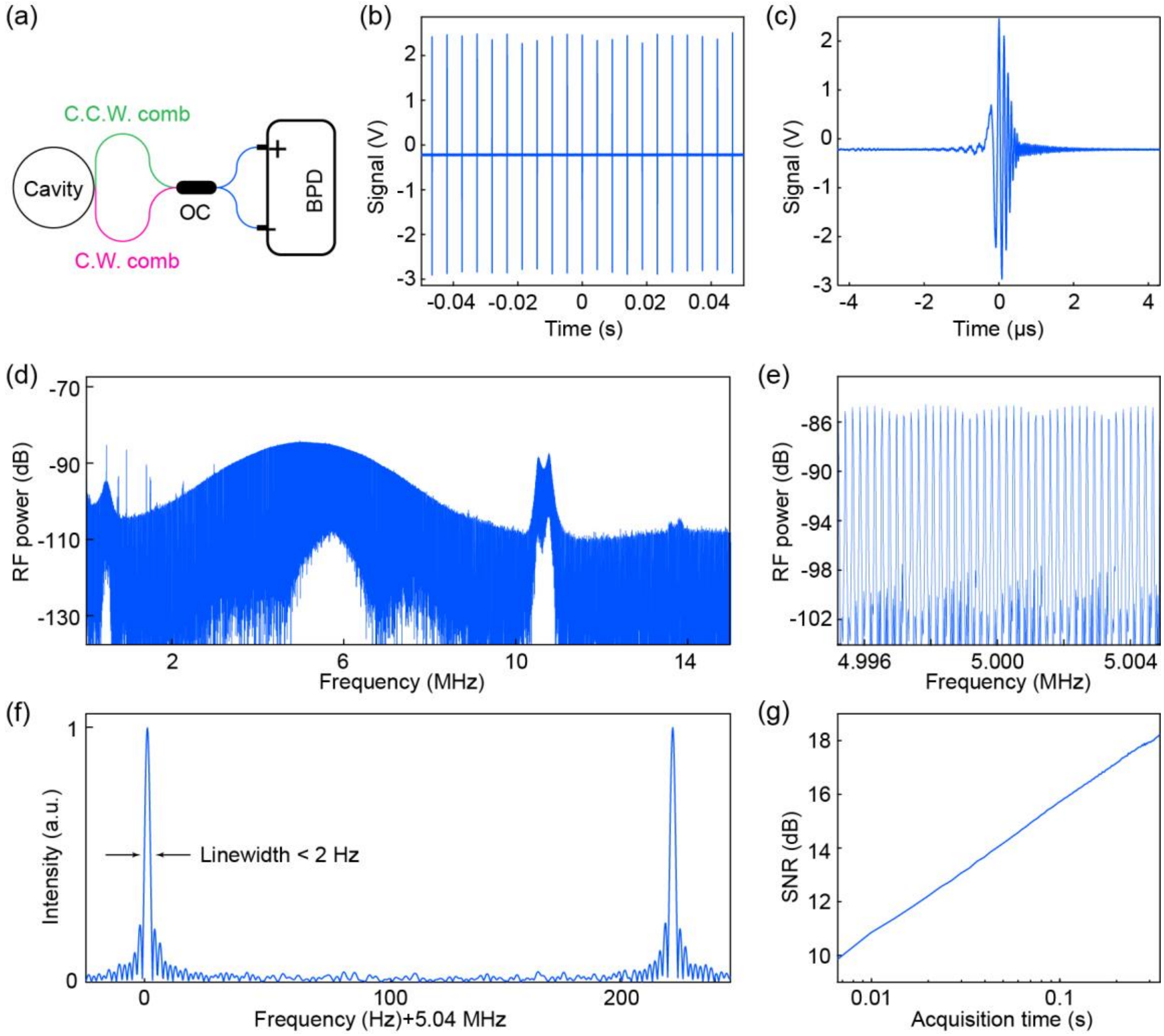

**Figure 3.** Free-running multi-heterodyne beat measurement. a) Experimental setup for detecting the multi-heterodyne beat signal. b) Time-domain interferograms generated by the two counter-propagating combs. c) Zoomed-in view of a single interferogram. d) RF dual-comb spectrum obtained by Fourier transforming the interferogram signal. e) Magnified view of the resolved RF comb lines near 5 MHz. f) Magnified RF comb lines after interferogram alignment, showing a linewidth below 2 Hz. g) SNR as a function of acquisition time after interferogram alignment. OC, optical coupler; BPD, balanced photodetector.

Next, we performed multi-heterodyne measurements using the free-running dual-comb laser. As shown in Figure 3(a), the two combs were combined using a 50:50 optical coupler, and the resulting beat signal was detected with a balanced photodetector (250 MHz bandwidth). The output of the detector was recorded by a 12-bit oscilloscope (HDO6104A, Teledyne LeCroy) at a sampling rate of 500 MSa/s. Figure 3(b) presents the measured time-domain interferograms, with a magnified view shown in Figure 3(c). The interferograms were Fourier transformed to reveal an RF dual-comb spectrum, as shown in Figure 3(d). Over the 0–11 MHz RF range, approximately 49,000 individual comb lines spaced by 223.5 Hz are resolved, corresponding to an optical detection bandwidth of 5.4 THz (or 44 nm at 1570 nm). A magnified view of the resolved comb lines is provided in Figure 3(e). With an extended measurement time window (>0.5 s), the RF comb lines exhibit a Fourier-transform-limited linewidth below 2 Hz, as shown in Figure 3(f), which compares favorably to previously reported free-running systems [30-34]. Moreover, the mean SNR of the RF comb lines increases as the square root of acquisition time (Figure 3(g)), indicating that the free-running dual combs retained a high degree of mutual coherence [7].

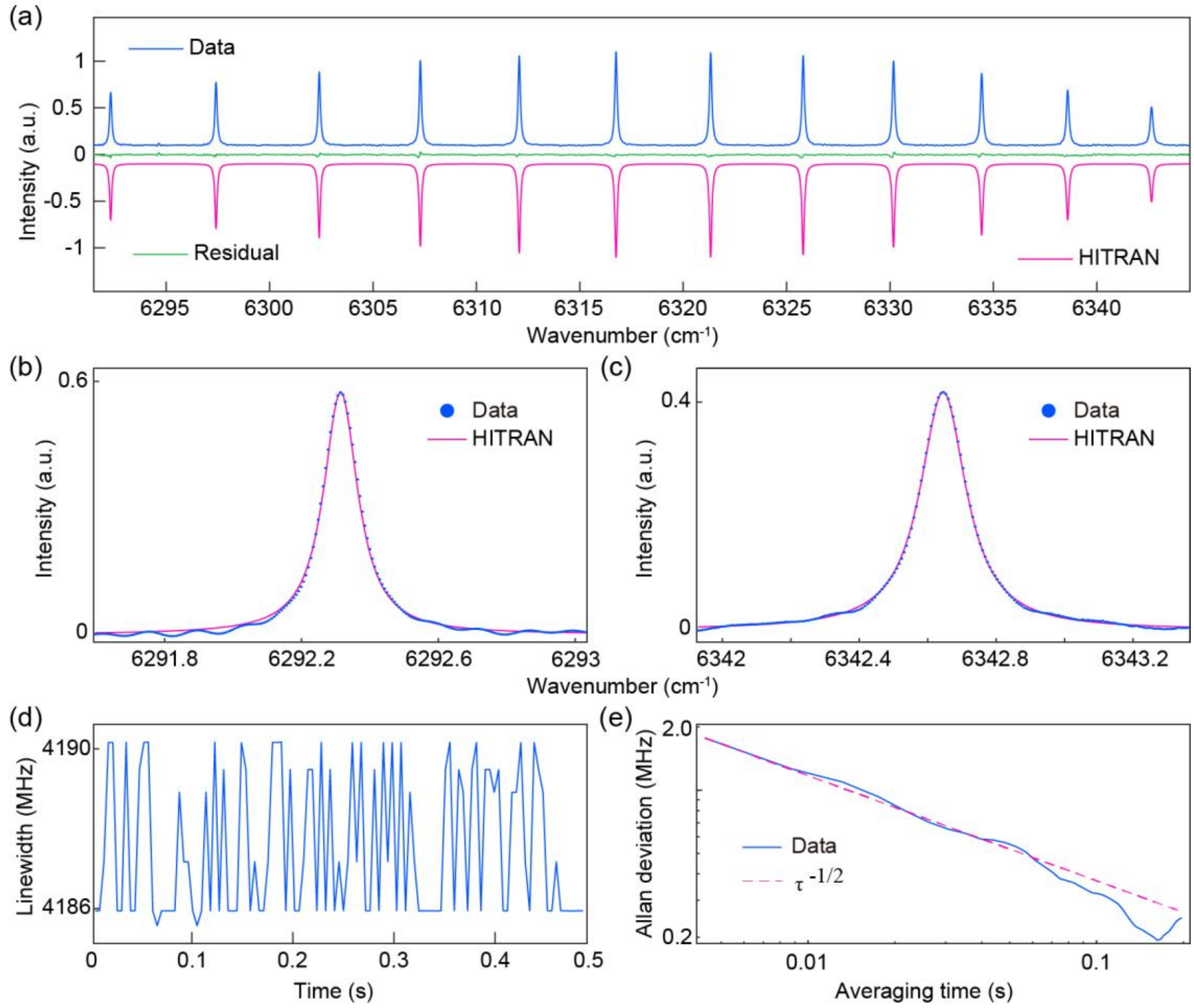

**Figure 4.** Gas absorption spectroscopy using the bidirectional SCDC system. a) Measured $^{12}CO$ P-branch absorption spectrum compared with the HITRAN-based simulation, with the residual shown in green. b, c) Zoomed-in views of the P(13) and P(2) absorption lines of the $^{12}CO$ (3-0) band. d) Time trace of the fitted linewidth for the CO absorption line at 6330.17 $cm^{-1}$, extracted from single-shot spectra. e) Allan deviation of the fitted linewidth as a function of averaging time.

Finally, we conducted dual-comb absorption spectroscopy on $^{12}CO$ gas. In this measurement, one of the two combs passed through a gas cell filled with a $^{12}CO$/air mixture at a total pressure of 1 atm, while the other comb served as the local oscillator. The measured spectrum was normalized by a gas-free background spectrum, and the wavenumber axis was calibrated using the center position of a known $^{12}CO$ absorption line at 6292.32 $cm^{-1}$. Figure 4(a) compares the measured rovibrational absorption lines in the $^{12}CO$ (3-0) band P-branch with a simulated spectrum calculated using the line parameters from the HITRAN2024 database [36]. The total measurement time was 500 ms. For clarity, the measured and simulated spectra are vertically offset. The measured spectrum agrees well with the simulations over the entire measured spectral range, with a residual SD below 0.17%. Figures 4(b) and 4(c) present zoomed-in views of two representative absorption lines. The measured spectra agree well with the simulations not only at the line centers but also across the spectral wings, confirming the capability of our dual-comb system for accurate molecular line-shape retrieval.

Moreover, our system enables real-time monitoring of spectroscopic parameters. As an example, we extracted the linewidth of the absorption feature at 6330.17 $cm^{-1}$ by fitting each single-shot spectrum, acquired within 4.4 ms, with a Voigt line-shape model. Figure 4(d) presents the time-dependent variations of the fitted linewidth, which remains centered at 4.19 GHz over the measurement period. Statistical analysis yields an Allan deviation of only 1.77 MHz, corresponding to ~$4.22 \times 10^{-4}$ of the pressure-broadened linewidth (4.2 GHz). The Allan deviation plot in Figure 4(e) further shows that the linewidth precision continues to improve with increasing averaging time. Note that with further improvement in absolute optical frequency calibration, our system holds potential for real-time molecular line position determination. Nevertheless, since molecular linewidths depend on pressure and temperature via line-shape broadening mechanisms, the current system is already well suited for applications such as thermometry and combustion diagnostics [37].

## 4. Discussion

Notably, conventional NPR-based bidirectional fiber lasers are vulnerable to environmentally induced birefringence variations, which can destabilize the mode-locked state and the relative repetition rate [38]. Our system overcomes this limitation and offers significantly enhanced frequency stability. Consequently, it supports broadband, comb-line-resolved DCS without active stabilization or phase correction, a key advantage for compact spectroscopic instruments requiring both broad coverage and simple implementation. It is also worth noting that dual-comb systems based on electro-optic modulators offer remarkable simplicity and excellent mutual coherence for comb-line-resolved spectroscopy [2-4]. However, their spectral bandwidth and the number of comb lines remain limited. In contrast, our dual-comb system resolves 49,000 individual comb lines across a 5.4 THz (44 nm) bandwidth.

Meanwhile, several aspects of the current implementation warrant further development. First, the present wavenumber calibration relies on known molecular absorption features. Applications requiring independent tracking of line-center shifts—such as Doppler-based flow velocimetry—would benefit from an external absolute-frequency reference or a more accurate real-time calibration strategy [39, 40]. One possible route is heterodyne referencing of the combs to narrow-linewidth continuous-wave lasers, enabling absolute frequency calibration and improved long-term accuracy. Second, the current average output power remains relatively low (~3 mW), limiting sensitivity and spectral extension. Power scaling through fiber amplification could increase per-line signal strength while simultaneously facilitating nonlinear spectral broadening. Combined with integrated nonlinear waveguides, such approaches may extend spectral coverage from the visible to the mid-infrared. In addition, optical-to-terahertz conversion using photoconductive antennas could further expand the platform toward high-precision THz spectroscopy [41, 42]. With these advances, the proposed dual-comb source could enable a broader range of line-shape-resolved molecular spectroscopy applications, including multispecies sensing, non-contact pressure and temperature diagnostics, combustion monitoring, and Doppler-based flow-velocity measurements [40, 43]. Beyond molecular spectroscopy, dual-comb techniques have also been extended to other fields, such as ranging and imaging [44-46], suggesting that this compact and highly coherent SCDC source may serve as a versatile platform for broader applications.

## 5. Conclusion

In summary, we have demonstrated a free-running, all-PM NPR mode-locked bidirectional SCDC fiber laser. This compact system generates two optical combs with a relative linewidth below 2 Hz. The free-running system exhibits significantly improved frequency stability compared to previous demonstrations. This capability enables comb-line-resolved, high-resolution molecular spectroscopy with broad applicability in gas sensing.

## Acknowledgements

This work is funded in part by Quantum Science and Technology-National Science and Technology Major Project (No. 2023ZD0301000).

## Data Availability Statement

The data that support the findings of this study are available from the corresponding author upon reasonable request.